\documentclass[12pt]{article}
\usepackage{amssymb,amsmath,esint,titlesec,mathrsfs}
\titleformat*{\section}{\normalsize\bf}
\titleformat*{\subsection}{\small\bf}

\begin{document}


\begin{titlepage}

\setlength{\baselineskip}{18pt}

                               \vspace*{0mm}

                             \begin{center}

{\large\bf Quantization in Cartesian coordinates and the Hofer metric}

                                   \vspace{20mm}

              \small\sf  NIKOLAOS \  KALOGEROPOULOS $^\ast$\\

                            \vspace{1mm}

                   {\small\sf  Center for Research and Applications \\
                                  of Nonlinear Systems  \   (CRANS),\\
                          University of Patras,
                                     Patras 26500, Greece.\\ }

                            \vspace{15mm}
                         
                 \small\sf CHRISTOS \  KOKORELIS $^\dagger$\\        
                           
                               \vspace{1mm}
                               
                {\small\sf  College of Data Science and Engineering,\\
                            The American University of Malta,\\  
                            Triq Dom Mintoff, Bormla BML 1013, Malta.\\  }                       
                         
                                    \end{center}

                            \vspace{10mm}

                     \centerline{\large\bf Abstract}
                                
                                     \vspace{3mm}
                     
    \noindent P.A.M. Dirac had stated that 
    the Cartesian coordinates are uniquely suited for expressing the canonical commutation relations in a simple form. 
    By contrast,  expressing these commutation relations in any other coordinate system is more complicated and less obvious. 
    The question that we address in this work, is the reason why this is true.
    We claim that this unique role of the Cartesian coordinates  is a result of the existence and uniqueness 
    of the Hofer metric on the space of canonical transformations of the phase space of the system 
    getting quantized.

                           \vspace{3mm}
                     
\normalsize\rm\setlength{\baselineskip}{18pt} 

                           \vfill


                             \vfill

\noindent\rule{12.5cm}{0.3mm}\\  
   \noindent   {\small\rm $^\ast$        Electronic mail: \ \  {\sf nikos.physikos@gmail.com}} \hspace{7mm}  {\small\rm (Corresponding author)}\\
    \noindent  {\small\rm $^\dagger$ Electronic mail: \ \  {\sf christos.kokorelis@aum.edu.mt}}\\
\end{titlepage}
 

                                                                                \newpage                 

\rm\normalsize
\setlength{\baselineskip}{18pt}

\section{Introduction}

In his foundational book on Quantum Mechanics \cite{Dirac} on the footnote on page 114, P.A.M. Dirac referring to the canonical commutation relations in quantizing Hamiltonian systems, 
states that \emph{``This assumption is found in practice to be successful only when applied with the dynamical coordinates and momenta referring to a Cartesian 
system of axes and not to more general curvilinear coordinates".} The statement points out the special role of the Cartesian coordinates on which the 
Heisenberg commutation relations have a simple form. However, the lesson learnt initially from Special and later from General Relativity, is that coordinates have no intrinsic physical meaning
but are just labels, representing the variables that one observer uses to express his physical measurements. The form of physical variables and laws cannot be dependent on any coordinates, 
expressed by stating that they should be ``covariant" under smooth coordinate re-parametrizations (diffeomorphisms).
 This is in stark contrast to the exalted status afforded on the Cartesian coordinates in canonical quantization which is pointed out in P.A.M. Dirac's statement. \\

The question which naturally arises, is ``Why?" Why should the Cartesian coordinates have this special role in canonical quantization, as opposed to any other coordinate system? \\

One can ignore this question, by just considering initially the commutation relations in the Cartesian coordinate system to be the staring point  from which one can convert them  to any other 
coordinate system and move on in quantizing such systems. This ``realistic approach" avoids even acknowledging  the question altogether. 
Another way to deal with this issue is to formulate the commutation relations in an explicitly coordinate-free way  as is done in 
Geometric Quantum Mechanics which can be considered as a part of Geometric Quantization \cite{GQ1, GQ2}. 
 A different approach, still coordinate independent but within the general framework of canonical quantization, 
is deformation quantization \cite{DQ1, DQ2, DQ3}. 
However these two approaches can  be seen in a far wider context as addressing other problems, such as manifest coordinate invariance and possibly operator ordering, 
associated to canonical quantization, rather than as aiming to address the issue we are considering.\\

One should also consider the resolution of this question in the context of the recently proposed, and closely related to Geometric Quantization,  the ``Brane Quantization" \cite{GW}.  
This approach however appears to be applicable to specific systems having very specific  symmetry groups, rather than being a prescription on how to quantize a generic system. 
Two additional proposals along the lines of canonical quantization are ``Enhanced Quantization" \cite{Kl1} and  ``Affine Quantization" \cite{Kl2, Kl3}. 
Topics related to phase space quantization and the major issue of the operator ordering problem (the Born-Jordan vs. Weyl ordering proposals) are discussed in \cite{deG1, deG2}. 
These and other related works  have pointed out the issue, directly or indirectly, of the special role of the Cartesian coordinates, without necessarily addressing it though, 
and therefore can be seen as providing the main starting point and impetus for the present investigation. \\     
 
 In an attempt to make the presentation self-contained, we provide some basic statements about the geometry of symplectic and Hamiltonian diffeomorphisms,  
 which even though are widely known in the Mathematics community, may not necessarily be familiar to our intended audience. Standard references in these topics are 
 \cite{Polt, McDS}. For most of this background, and where no references are provided,  we follow closely and extensively \cite{McDS} and references therein. 
 In Section 2, we make some general comments about Riemannian metrics 
 and introduce the Hofer metric on the space of compactly supported Hamiltonian diffeomorphisms. In Section 3, we indicate how the uniqueness of the Hofer metric, 
 among all \ $L^p$ \ metrics on this space singles out the Cartesian among all other coordinates in canonical quantization. Some general comments and an assessment is 
 presented in Section 4. \\  
 
 
\section{Some symplectic background} 
 
 \subsection{Notation}
 
To address this question we follow the spirit of the approach of P.A.M. Dirac. We start, and to set the notation, 
 by considering a $2n$-dimensional symplectic manifold \ $(\mathcal{M}, \omega)$ \ which 
represents the phase space \  $\mathcal{M}$ \ and also encodes the Hamiltonian evolution of the system under study through \ $\omega$ \  in a coordinate-free way. 
Following the standard notation, let \ $T\mathcal{M}$ \ indicate the tangent bundle, and \ $T^\ast\mathcal{M}$ \ the cotangent bundle of \ $\mathcal{M}$ \ and \ $\mathsf{Diff}(\mathcal{M})$ 
\ the group of diffeomorphisms of \ $\mathcal{M}$. \ 
Since regularity of maps will not be of any interest in this work, despite its considerable role in Symplectic Geometry, 
we will assume in the sequel that all functions, vector fields etc are sufficiently smooth for our purposes.
Let \ $\mathcal{M}$ \ be locally parametrized by \ $(q_1, \ldots q_n, p_1, \ldots p_n)$ \ where the pairs \ \ $(q_i, p_i), \ i=1, \ldots , n$ \  \ are made up of canonically conjugate 
variables with respect to each other.  This, as is well-known, means that if \ $X_f\in T\mathcal{M}$ \ indicates the vector field corresponding  to a function 
\  $f: \mathcal{M} \longrightarrow \mathbb{R}$ \ through
\begin{equation} 
    \omega (X_f, \cdot) \ = \ df
\end{equation}
then the Poisson bracket between two functions \ $f, g: \mathcal{M} \longrightarrow \mathbb{R}$ \ is defined by 
\begin{equation}
     \{f, g\} \ = \ \omega (X_f, X_g) \ = \ df(X_g)
\end{equation}
and one finds for the canonical variables 
\begin{equation}
           \{q_i, p_j\} = \delta_{ij}, \hspace{10mm} \{q_i, q_j \} = 0, \hspace{10mm} \{p_i, p_j\} = 0, \hspace{10mm} i,j = 1,\ldots, n
\end{equation}
as is well-known. Vector fields \ $X_f$ \ obeying (1) are called Hamiltonian vector fields. It should be noted at this point that, even though this terminology is motivated by Classical Mechanics,  
$f$ does not have to be the actual Hamiltonian of a physical system, but it can just be any sufficiently smooth real-valued function on \ $\mathcal{M}$.  \\


\subsection{Riemannian metrics on symplectic manifolds} 

We digress in this paragraph to address an occasional misconception. 
It is indeed true that the symplectic manifold \ $(\mathcal{M}, \omega)$ \ does not possess a ``natural" metric \cite{Kl1, Kl2, Kl3}.  However, in a symplectic manifold, 
there are almost complex structures \ $\mathbb{J}$, \ namely automorphisms of \ $T\mathcal{M}$,  \ 
\begin{equation}
 \mathbb{J}: T\mathcal{M} \longrightarrow T\mathcal{M}, \hspace{15mm} \mathbb{J}^2 = -\mathbf{1}  
\end{equation}
integrable or not, which have the following two properties: 
\begin{itemize}
     \item they are ``tamed" by \ $\omega$, \ namely
                    \begin{equation}
                                        \omega (Y, \mathbb{J}Y) \ \geq \  0, \hspace{15mm} \forall \ Y \in T\mathcal{M}\backslash \{ 0 \}
                    \end{equation}
      \item they leave \ $\omega$ \ invariant, namely 
                    \begin{equation}               
                                   \omega (\mathbb{J}Y, \mathbb{J}Z) \  =  \  \omega(Y, Z), \hspace{10mm} \forall \ Y,Z \in T\mathcal{M}
                    \end{equation}               
\end{itemize}
In (4),  \ $\mathbf{1}$ \ stands for the identity map. 
Almost complex structures \ $\mathbb{J}$ \ having both of the above properties are called ``compatible" with the the symplectic structure \ $\omega$. \ 
Then the bilinear form defined by 
\begin{equation}
       \mathbf{g} (Y,Z) \ = \ \omega (Y, \mathbb{J}Z)
\end{equation}
is a Riemannian metric on \ $\mathcal{M}$ \cite{McDS}. It turns out that the space of compatible with \ $\omega$ \ almost complex structures \ $\mathbb{J}$ \ 
on a symplectic manifold \ $\mathcal{M}$ \ is convex and therefore contractible. 
So, the issue is not so much that a Riemannian metric does not exist on a phase space $\mathcal{M}$, but that there are too many such Riemannian metrics, 
and the use of each one of them  may give rise to different, and therefore incompatible with the others,  physical predictions. 
As a result, the problem shifts to understanding which one of these metrics, if any, can be used to extract the relevant 
physical information from the model under study. All such metrics, being bilinear forms,  
fit most simply and naturally in arguments involving either real or complex-valued square integrable functions (functions  of Lebesgue integrability class \ $L^2(\mathcal{M})$).\\


\subsection{Symplectic and Hamiltonian groups of maps}

If the Riemannian metrics on symplectic manifolds are most ``natural", based on the above, one can ask the same question for the space of symplectic diffeomorphisms 
(``symplectomorphisms") of symplectic manifolds \ $\mathcal{M}$. \ This in the spirit of F. Klein's ``Erlangen program" where geometric properties of a space 
can be deduced from the set of  its symmetries. 
Symplectomorphisms are  diffeomorphisms \ $\psi: \mathcal{M} \longrightarrow \mathcal{M}$ \ which preserve the symplectic 
structure, namely
\begin{equation} 
                    \psi^\ast (\omega) \  = \  \omega
\end{equation}
Symplectomorphisms are, somewhat inaccurately stated,  the ``canonical transformations" of Classical Mechanics.
Let us indicate by \  $\mathsf{Symp}(\mathcal{M}, \omega)$ \  the set of symplectomorphisms of \ $(\mathcal{M}, \omega)$, \ namely
\begin{equation}
              \mathsf{Symp}(\mathcal{M}, \omega) \ = \ \{ \psi \in \mathsf{Diff}(\mathcal{M}) : \psi^\ast(\omega) = \omega \}
\end{equation} 
This set becomes a group under the composition of maps and it is a subset of \ $\mathsf{Diff}(\mathcal{M})$. \  It is true that \ $\mathsf{Symp}(\mathcal{M})$ \ 
is an infinite dimensional group, something that is intimately related to the fact that all symplectic manifolds of dimension \ $2n$ \ are locally diffeomorphic to 
\ $(\mathbb{R}^{2n}, \omega_0)$ \ according to Darboux's theorem, where 
\begin{equation} 
   \omega_0 \ = \ \sum_{i=1}^n dq_i \wedge dp_i
\end{equation}
The situation is in sharp contrast to the Riemannian manifold one where the isometry groups are finite dimensional. In a more algebraic language, someone can say that 
these form a category where the objects are symplectic manifolds and the morphisms are the symplectomorphisms. The category-theoretic language has proved to be 
quite fruitful  in some lines of investigation, although we will not pursue it further in this work.\\

To proceed (we closely follow \cite{McDS} in the rest of this and in the next subsection), let us assume that \ $(\mathcal{M}, \omega)$ \ is a connected symplectic manifold without boundary,
which does not have to be compact. A symplectomorphism \ $\psi: \mathcal{M} \longrightarrow \mathcal{M}$ \ is called ``compactly supported" if there exists a 
compact subset \ $U\subset \mathcal{M}$ \ such that \ $\psi\big|_{\mathcal{M}\backslash U} = \mathbf{1}$. \ As in the general case, one can see that the set of 
compactly supported symplectomorphisms of \ $(\mathcal{M}, \omega)$ \ also forms a group, indicated by \ $\mathsf{Symp_c}(\mathcal{M}, \omega)$. \ Compact 
support is used to express the fact that a map only acts non-trivially within a compact set, and acts trivially outside it.  It is a form of ``localization" of the domain of the 
maps under study and is used extensively in Physics, even when not explicitly stated. \\

What one needs, and is extensively used in Classical Mechanics as the effective definition of canonical transformations, is a set of maps slightly less general than the symplectomorphisms. 
To begin with, a symplectic isotopy of \ $(\mathcal{M}, \omega)$ \ is a smooth map 
\begin{equation}
\psi: [0,1] \times \mathcal{M} \longrightarrow \mathcal{M}, \  \hspace{15mm} (t, x) \mapsto \psi_t(x),  \hspace{15mm} \forall \ t\in [0,1], \hspace{2mm} \forall \ x\in\mathcal{M}  
\end{equation}
such that \ $\psi_t$ \ is a symplectomorphism for all \ $t\in [0,1]$ \ and \ $\psi_{t=0} = \mathbf{1}$.  \  Any such symplectic isotopy is generated by a family of vector fields 
\ $X_t\in T\mathcal{M}$ \ by differentiation, namely 
\begin{equation}
   \frac{d}{dt} \psi_t \ = \ X_t, \hspace{15mm} \psi_0 \ = \ \mathbf{1}
\end{equation}
All such vector field \ $X_t$ \ are symplectic, namely they obey 
\begin{equation}
     d \omega (X_t, \cdot) \ = \ 0
\end{equation}
The 1-form \ $\omega(X_t, \cdot)$ \ is closed according to (13). The transition from ``symplectic" to ``Hamiltonian" takes places by requiring such forms to be exact. 
To be concrete, a symplectic isotopy \ $\psi_t, \ t\in[0,1]$ \ is called a Hamiltonian isotopy, if the 1-form \ $\omega(X_t, \cdot)$ \ is exact. Then the corresponding vector field 
\ $X_t$, \  is called a Hamiltonian  vector field. when this is true then a version of (1) is satisfied, namely, there is a function \ 
$H: [0,1] \times\mathcal{M} \longrightarrow \mathcal{M}$ \ such that 
 \begin{equation}
       \omega (X_t, \cdot) \ = \ dH_t, \hspace{15mm} \forall \ t\in [0,1]
\end{equation} 
In accordance with (1), this function is called a ``time-dependent (non-autonomous) Hamiltonian" 
and is determined by the Hamiltonian isotopy only up to a ``time-dependent" additive function \ $c: [0,1] \rightarrow \mathbb{R}$. \  A symplectomorphism \ 
$\psi \in \mathsf{Symp}(\mathcal{M}, \omega)$ \  is called a  Hamiltonian symplectomorphism, if there is a Hamiltonian isotopy \ $\psi_t \in \mathsf{Symp}(\mathcal{M}, \omega)$ \ 
from \ $\psi_0 = \mathbf{1}$ \  to \ $\psi_1 = \psi$. \ Let the space of Hamiltonian symplectomorphisms of \ $(\mathcal{M}, \omega)$ \ be indicated by \ $\mathsf{Ham}(\mathcal{M}, \omega)$. \ 
It turns out that \ $\mathsf{Ham}(\mathcal{M}, \omega)$ \ is a path-connected normal subgroup of \ $\mathsf{Symp}(\mathcal{M, \omega})$. \ Moreover, if \ $\mathcal{M}$ \ is compact and 
without a boundary, then \ $\mathsf{Ham}(\mathcal{M}, \omega)$ \ is simple, namely it contains no nontrivial normal subgroups \cite{Ban}.  \\

It is also true, but equally nontrivial, that
the Lie algebra of  \ $\mathsf{Ham}(\mathcal{M}, \omega)$ \  is the Lie algebra of all Hamiltonian vector fields. 
A property that makes all such Hamiltonian vector fields highly desirable, is that their action preserves the symplectic form: 
the Lie derivative \ $D_H$ \  of the symplectic form along a Hamiltonian vector field  is zero. Indeed, using Cartan's formula
\begin{equation} 
      D_H \omega \ = \ (di_H +i_Hd)\omega \ = \ d(\omega(X_H, \cdot)) + i_H (d\omega) \ = \ ddH \ = \ 0 
\end{equation}
where \ $i_H$ \ indicates the contraction along the Hamiltonian vector field \ $X_H$, \ we have used the fact that \ $\omega$ \ is closed, by definition, and (14), 
where \ $H$ \ is the Hamiltonian function generating the Hamiltonian vector field \ $X_H$.\\

Since we mentioned the above groups and algebras as being ``Lie",
the latter having a topological structure, we have to mention, for completeness, that this group is tacitly assumed to be  endowed with the  \ $C^\infty$ \ 
compact-open topology, which is also intuitively ``natural" and quite familiar  in various  physical applications. To be more precise, it turns out that \ 
$\mathsf{Ham}(\mathcal{M}, \omega)$ \ is an infinite dimensional  Fr\'{e}chet Lie group endowed with the \ $C^\infty$ \ compact-open topology. \\

Combining the above concepts, one can see that every Hamiltonian function \ $H: [0,1]\times \mathcal{M} \rightarrow \mathbb{R}$ \ having compact support, determines a 
Hamiltonian isotopy with compact support \ $\psi_t, \ t\in [0,1]$ \ via (12), and the  time one diffeomorphism \ $\phi_H = \psi_1$  \ is called compactly supported Hamiltonian symplectomorphism.
Let \ $\mathsf{Ham}_c(\mathcal{M}, \omega)$ \ indicate the set of compactly supported Hamiltonian symplectomorphisms, namely 
\begin{equation}
    \mathsf{Ham}_c (\mathcal{M}, \omega) \ = \ \left\{ \phi_H : H \in C_0^\infty ([0,1] \times \mathcal{M}) \right\}
\end{equation}
where \ $C_0^\infty ([0,1] \times\mathcal{M})$ \ indicates the set of infinitely continuously differentiable (``smooth") functions with compact support on the set \ $[0,1]\times \mathcal{M}$. \ 
It should be obvious that if \ $\mathcal{M}$ \ is compact and without boundary,  then \ $\mathsf{Ham}_c(\mathcal{M}, \omega)  =  \mathsf{Ham}(\mathcal{M}, \omega)$. \\ 

Before closing this subsection it may be worth pointing out, mostly for completeness, the relation between the Symplectic and Hamiltonian groups of maps. 
Let \ $\mathsf{Symp}_0(\mathcal{M}, \omega)$ \ indicate the set of symplectomorphisms of \ $(\mathcal{M}, \omega)$ \ which are connected to the identity. This means that 
\  $\psi\in \mathsf{Symp}(\mathcal{M}, \omega)$ \ if there is a symplectic isotopy \ $\psi_t, \ t\in [0,1]$ \ from the identity \ $\mathbf{1}$ \ to \ $\psi$. \  Then clearly \ 
$\mathsf{Ham}(\mathcal{M}, \omega) \subset \mathsf{Symp}_0(\mathcal{M}, \omega)$. \ The relation between the set of Symplectic and Hamiltonian diffeomorphisms 
essentially reflects the relation between closed and exact 2-forms, so it is homological. To be more specific, there is an exact sequence 
\begin{equation}      
        0 \  \longrightarrow \  \mathsf{Ham}(\mathcal{M}, \omega) \  \longrightarrow \   \mathsf{Symp}_0(\mathcal{M}, \omega) \   \longrightarrow \  H^1(\mathcal{M}, \mathbb{R})/\Gamma \
              \longrightarrow \  0
\end{equation}
where $H^1(\mathcal{M}, \mathbb{R})$ is the first singular cohomology group of \ $\mathcal{M}$ \ with real coefficients, and \ $\Gamma$ \ is a countable subgroup of it. \
$\Gamma$ \  is called the flux group of \ ($\mathcal{M}, \omega$) \ and is the fundamental group of the  group \ $\mathsf{Symp}_0(\mathcal{M}, \omega)$ \ under the flux homomorphism
\cite{McDS}. Since we have no actual need for this and related developments in this work, we will not elaborate upon them any further.\\


\subsection{The Hofer metric}

The question of the existence and explicit construction of metrics on infinite dimensional manifolds, even if they are Lie groups, is a non-trivial one. We state the ``even if" since 
finite-dimensional Lie groups are not only useful in Physics, but have very nice mathematical properties which has allowed their classification. 
One of the crucial properties of such finite-dimensional groups is the relation between the linear
structure provided by the Lie algebra of left invariant vector fields at the origin, and the whole non-linear structure of the group, which is getting implemented via the exponential map. 
By contrast,  infinite dimensional Lie groups are quite different \cite{Milnor, Neeb}. There is no guarantee that the exponential map exists, and even if it does, it does not have to be surjective. 
And all this after ignoring issues related to the underlying topologies, or the model spaces upon which these groups are built, which play a crucial role in infinite dimensions. 
It is therefore surprising when a metric is found on such infinite dimensional Lie groups and even more so when it has nice equivariant properties, and especially if it is bi-invariant. \\

For a finite-dimensional Lie group \ $\mathcal{G}$ \ with corresponding Lie algebra  \ $\mathfrak{g}$ \ a norm $\| \cdot \|$ on \ $\mathfrak{g}$ \  is called invariant, if it is invariant under the adjoint 
action \  $\mathcal{G} \times \mathfrak{g} \longrightarrow \mathfrak{g}$ \  given by 
\begin{equation}    
           \mathfrak{g} \ni X \   \longmapsto \  g^{-1}Xg \in \mathfrak{g}, \hspace{15mm}  \forall  \  X\in \mathfrak{g}, \  \ \  \forall \ g \in \mathcal{G}
\end{equation}
Such a  norm induces a (Finslerian) distance function \ $\mathsf{d}: \mathcal{G}\times\mathcal{G} \rightarrow \mathbb{R}_+$ \  given by 
\begin{equation}
          \mathsf{d}(g_0, g_1) \ = \ \inf_{g\in\mathcal{G}}  \   \int_0^1 \Big|\!\Big| \frac{dg(t)}{dt} g^{-1}(t) \Big|\!\Big| \ dt      
\end{equation}
where the infimum is taken over all paths \ $g: [0,1] \times \mathcal{G} \longrightarrow \mathcal{G}$ \ connecting \ $g_0 = g(0)$ \ and \ $g_1 = g(1)$.  \
By analogy, the Hofer norm \cite{Hofer} on the Lie algebra of compactly supported Hamiltonian functions is defined to be the ``oscillation norm" 
\begin{equation}  
      \big|\!\big| H \big|\!\big| \ = \ \max_\mathcal{M} H - \min_\mathcal{M} H
\end{equation}
for \ $H \in C^\infty_0 (\mathcal{M})$. \ In case \ $\mathcal{M}$ \  is compact, the constant functions have zero norm, therefore (20) defines a norm on the 
quotient space \ $C^\infty(\mathcal{M})/\mathbb{R}$ \ or, equivalently, on the space of Hamiltonian vector fields. The length functional on the space of 
Hamitonian isotopies \ $\{ \psi_t \}, \  t\in [0,1]$ \  with compact support in  \ $[0,1]\times \mathcal{M}$ \ is     
\begin{equation}
     l_H(\{ \psi \}) \ = \ \int_0^1 \left( \sup_{\mathcal{M}} H_t - \inf_{\mathcal{M}} H_t\right) \ dt 
\end{equation}
where \ $H_t$ \  is the Hamiltonian function with compact support generating the isotopy \ $\psi_t$. \  As a result of (21), one can define the Hofer distance \ 
$\mathsf{d}_H (\phi_1, \phi_2)$ \ between two compactly supported Hamiltonian symplectomorphisms \ $\phi_1, \phi_2 \in \mathsf{Ham}_c(\mathcal{M}, \omega)$ \ 
as the infimum of the lengths of the Hamiltonian isotopies with compact support joining them, namely   
\begin{equation} 
       \mathsf{d}_H (\phi_1, \phi_2) \ = \ \inf_{\phi_H} \int_0^1 \big|\!\big| H_t \big|\!\big| \ dt 
\end{equation}
where 
\begin{equation} 
        \phi_H \ = \  \phi_2^{-1} \circ \phi_1 
\end{equation}
is the infimum taken over all compactly supported Hamiltonian functions \ $H$ \ generating the Hamiltonian symplectomorphism (23). It is straightforward to prove that 
the Hofer distance (22) is symmetric
\begin{equation}
      \mathsf{d}_H (\phi_1, \phi_2) \ = \ \mathsf{d}_H (\phi_2, \phi_1), \hspace{15mm} \forall \ \phi_1, \phi_2 \in \mathsf{Ham}_c(\mathcal{M}, \omega)  
\end{equation}
satisfies the triangle inequality 
\begin{equation}
   \mathsf{d}_H (\phi_1, \phi_2) \ \leq \ \mathsf{d}_H (\phi_1, \phi_3) + \mathsf{d}_H (\phi_3, \phi_2), \hspace{15mm}
                                                                                          \forall \ \phi_1, \phi_2, \phi_3 \in \mathsf{Ham}_c(\mathcal{M}, \omega) 
\end{equation}
and that it is bi-invariant 
\begin{equation}
    \mathsf{d}_H (\phi_1 \circ \phi_0, \phi_2 \circ \phi_0) \ = \ \mathsf{d}_H (\phi_0 \circ \phi_1, \phi_0 \circ \phi_2) \ = \ \mathsf{d}_H (\phi_1, \phi_2), \hspace{3mm}
                 \forall \ \phi_0, \phi_1, \phi_2 \in \mathsf{Ham}_c (\mathcal{M}, \omega) 
\end{equation}
and
\begin{equation}
  \mathsf{d}_H (\phi_0^{-1} \circ \phi_1 \circ \phi_0, \phi_0^{-1} \circ \phi_2 \circ \phi_0)  =  \mathsf{d}_H(\phi_1, \phi_2),  \hspace{3mm} 
                                         \forall \ \phi_1, \phi_2 \in \mathsf{Ham}_c(\mathcal{M}, \omega),  \ \forall \ \phi_0 \in \mathsf{Symp}_c(\mathcal{M}, \omega)  
\end{equation}
By contrast, it is quite non-trivial to prove that the Hofer distance is non-degenerate \cite{Hofer, LMcD},
\begin{equation}
       \mathsf{d}_H (\phi_1, \phi_2) \ = \ 0 \hspace{5mm} \implies \hspace{5mm} \phi_1 \ = \ \phi_2
\end{equation}
 therefore it defines a legitimate metric on \ $\mathsf{Ham}_c(\mathcal{M}, \omega)$. \ It should come as no surprise that the explicit computation of the Hofer 
 distance between any two elements of \ $\mathsf{Ham}_c(\mathcal{M}, \omega)$  \ is a hard task, practically impossible to explicitly perform except in very special cases.\\  


\subsection{Alternative formulations of the Hofer metric}

A slight alternative to the above definition of the Hofer metric is related to the way one defines the space of maps giving rise to the Hamiltonian diffeomorphisms \cite{Polt}. 
The reason for this definition of the functional space is two-fold, at least: in the case that \ $\mathcal{M}$ \ is an open manifold, namely a non-compact manifold without boundary, 
as is frequently the case in Physics, makes the Hamiltonian vector fields to be compactly supported. This eliminates the possibility of  having Hamilton's  evolution equations 
blow up at finite time. Second, definitions of maps based such a functional space eliminates the freedom that one has to add some arbitrary ``time-dependent" function \ 
$c: [0,1]\longrightarrow \mathbb{R}$ \ to the Hamiltonian, as was pointed out in the discussion following (14). Consider the set of smooth functions on \ $\mathcal{M}$ \ which are 
normalized so that that mean value is zero, with respect to the Liouville volume form on \ $\mathcal{M}$, \ namely consider the set \ $f: \mathcal{M} \longrightarrow \mathbb{R}$ \ such that 
\begin{equation}    
                   \int_\mathcal{M} f \  \frac{\omega^n}{n!} \ = \ 0 
\end{equation}   
This set forms an algebra under composition, which turns out to be the algebra of the Hamiltonian vector fields \ $\mathsf{ham}(\mathcal{M}, \omega)$ \ on \ $\mathcal{M}$. \ 
The rest of the construction of the Hofer metric proceeds in the same way as in the previous section.\\   

A second alternative formulation, actually a generalization,  of the Hofer metric was provided in \cite{Che}. 
To formulate this version of the Hofer metric, we need the concept of Lagrangian submanifolds of symplectic manifolds. 
A Lagrangian submanifold \ $\mathcal{L}$, \ of a $2n$-dimensional symplectic manifold \ $(\mathcal{M}, \omega)$, \ is a submanifold \ $\mathcal{M} \subset \mathcal{M}$ \ 
at which the restriction of the symplectic form \ $\omega$ \ is vanishing, namely 
\begin{equation}  
         \omega \big|_\mathcal{L} \ = \ 0
\end{equation}
and which has maximal possible dimension, which turns out to be 
\begin{equation}   
        \dim \mathcal{L} \ = \  \frac{1}{2} \dim \mathcal{M} \ = \  n
\end{equation}
This definition, despite its mathematical elegance and naturality, may appear to be too abstract to a Physicist. However, things become clearer if one thinks about
Lagrangian submanifolds as coordinate independent generalizations of either the  space of generalized positions \ $\{q_i\}, \ i=1,\ldots, n$ \ or the space of their canonical momenta 
\ $\{ p_i \}, \ i=1,\ldots, n$ \ of the phase space \ $\mathcal{M}$, \ in the case that \ $\mathcal{M}$ \ is the cotangent bundle of the manifold of generalized positions.  
The fact that this mental picture, physically motivated, is essentially correct is provided by A. Weinstein's Lagrangian neighborhood theorem \cite{Wein, McDS} which states, that 
if \ $\mathcal{L}$ \ is a compact Lagrangian submanifold of the symplectic manifold \ $(\mathcal{M}, \omega)$ \ then, there exists a neighborhood \ 
$N(\mathcal{L}_0)\subset T^\ast\mathcal{L}$ \ of the zero section \ $\mathcal{L}_0$, \ a neighborhood \ $\mathcal{U} \subset \mathcal{M}$, \ and a diffeomorphism \ 
$\phi: N(\mathcal{L}_0) \longrightarrow \mathcal{U}$  \ such that 
\begin{equation}   
      \phi^\ast\omega = -d\lambda, \hspace{15mm}  \phi\big|_{\mathcal{L}} = \mathbf{1}
\end{equation}
where 
\begin{equation} 
      \lambda \ = \ \sum_{i=1}^n \ p_i \ dq_i
\end{equation}
This theorem essentially states that the symplectomorphism class of a small neighborhood of the Lagrangian manifold \ $\mathcal{L}$ \ is completely determined by the diffeomorphism class 
of \ $\mathcal{L}$ \ itself. \\

Lagrangian submanifolds are objects of fundamental importance in symplectic geometry for the following reason:  let \ $(\mathcal{M}_1, \omega_1)$ \ and \ 
$(\mathcal{M}_2, \omega_2)$ \ be $2n$-dimensional symplectic manifolds.  Let \ $f: \mathcal{M}_1 \longrightarrow \mathcal{M}_2$  \ be  diffeomorphism, whose graph \ $\mathbf{\Gamma}_f$ \ 
is defined as
\begin{equation}
      \mathbf{\Gamma}_f \ = \ \left\{ (x, f(x)) : \  x\in\mathcal{M}_1 \right\}
\end{equation}  
Then \ $f$ \ is a symplectomorphism if and only if \ $\mathbf{\Gamma}_f$ \ is a Lagrangian submanifold of \ $\mathcal{M}_1 \times \mathcal{M}_2$ \ endowed with the symplectic form  \    
$\tilde{\omega} = \omega_1 \oplus (-\omega_2)$. \ This relation between symplectic diffeomorphisms and Lagrangian submanifolds makes Yu. Chekanov's generalization of the Hofer distance 
particularly useful. From a physicist's viewpoint and in the confines of the present work,  Lagrangian submanifolds are extensively used in Geometric Quantization, 
especially when someone chooses a polarization, and also in a similar approach followed in Brane Quantization.\\

Let \ $(\mathcal{M}, \omega)$ \ be a compact and without boundary symplectic manifold, and let \ $\mathcal{L}$ \ be a Lagrangian submanifold. Consider a family of Lagrangian submanifolds  
\  $\mathfrak{L}_t, \ t\in [0,1]$ \ such that each member of this family is diffeomorphic to \ $\mathcal{L}$. \ A path \  $\gamma$ \ connecting \ $\mathfrak{L}_0$ \ to \ $\mathfrak{L}_1$ \ is called exact
if there is a smooth map \ $\phi: [0,1] \times \mathcal{L}$ \ such that \ $\phi (t, \mathcal{L}) = \mathfrak{L}_t$ \ and \ $\phi^\ast \omega = dt \wedge dH(t)$ \ for a function \ 
$H: [0,1] \times \mathcal{L} \rightarrow \mathbb{R}$.  \ The length of such an exact path \ $\gamma$ \ is given by
 \begin{equation} 
      l_C(\gamma) \ = \ \int_0^1 \left\{ \max_{x\in\mathcal{L}} H_t(x) - \min_{x\in\mathcal{L}} H_t(x)  \right\}  \ dt
\end{equation}   
Let \ $\mathbf{L}(\mathcal{L}, \mathcal{M})$ \ be the set of all Lagrangian submanifolds of \ $\mathcal{M}$ \ which can be connected to \ $\mathcal{L}$ \ through an exact path.  Consider 
two Lagrangian submanifolds \ $\mathcal{L}_1, \ \mathcal{L}_2 \in \mathbf{L}(\mathcal{L}, \mathcal{M})$. \ We define the Hofer distance between \ $\mathcal{L}_1$ \ and \ $\mathcal{L}_2$  \ by 
\begin{equation}
             \mathsf{d}_C \ = \ \inf_\gamma l_C(\gamma)
\end{equation}
where the infimum is taken over all exact paths on \ $\mathbf{L}(\mathcal{L}, \mathcal{M})$ \  which connect \  $\mathcal{L}_1$ \ and \ $\mathcal{L}_2$. \ 
This is the analogue of the Hofer distance for the case of any two such Lagrangian submanifolds.
A comparison of Hofer's metric on the space of compactly supported Hamiltonian symplectomorphisms 
versus case of Lagrangian submanifolds  has been provided in \cite{Ost}.\\


\subsection{On the uniqueness of the Hofer metric}

The question about the uniqueness of the Hofer metric was initiated in \cite{EP}. Of course the word ``uniqueness" has to also be qualified by providing the 
set of possible alternatives. Since Hofer's metric (20), (22)  is of Lebesgue integrability class  \ $L^\infty$, \ the most obvious alternatives are provided by the $p$-norms used  
to define the other classical \ $L^p$ \ spaces, with \ $p\geq 1$, \ namely for distance functions \ $\mathsf{d}_p$ \ on \ $\mathsf{Ham}_c (\mathcal{M}, \omega)$ \ defined by
\begin{equation}
       \mathsf{d}_p (\phi_1, \phi_2) \ = \ \inf_{\phi_H} \  \int_0^1 \left( \int_\mathcal{M} |H|^p \  \frac{\omega^n}{n!}  \right)^\frac{1}{p} \ dt
\end{equation}
where \ $\phi_H$ \  is as in (23). It was proved in \cite{EP} that these distance functions are only giving pseudo-distances, rather than distances,  
as they do not satisfy the analogue of  the non-degeneracy condition (28). Things are far worse though with this alternative: it was also proved in \cite{EP}
that for compact symplectic manifolds \  $\mathcal{M}$ \ without boundary, the pseudo-distances  (37) vanish identically.  Therefore this leaves the Hofer metric 
as the only non-trivial bi-invariant Finsler metric on \ $\mathsf{Ham}(\mathcal{M}, \omega)$ \  for closed (namely, compact without boundary) manifolds \ $\mathcal{M}$. \\

Someone might wish to construct metrics on \ $\mathsf{Ham}(\mathcal{M}, \omega)$ \ or on \ $\mathsf{Symp}(\mathcal{M}, \omega)$ \ of other functional forms, 
not induced by any of the classical $L^p$-norms. 
An immediate generalization, for instance, motivated by recent advances in functional analysis and its related aspects of convex geometry, 
would be to consider more general Banach norms like the Orlicz/Luxemburg norms etc. However, some of these norms, even though of considerable interest in
several branches of Mathematics, do not appear to be ``too different"  from the \ $L^p$ \ ones, from a physicist's viewpoint,  to offer any radically new insights.    
To strengthen this viewpoint, one can notice that  canonical quantization of a system involves quantizing polynomials of the coordinates  
of the classical phase space of a system, in the vast majority of cases. For such cases, \ $L^p$ \ integrability of the associated states and  expressions involving
operator products is more than sufficient, as such states are vectors on Hilbert spaces, hence they involve square integrable spaces functions, or their dual spaces of 
(tempered) distributions. This does not mean that efforts toward constructing different norms from the \ $L^p$ \ ones on the spaces of Hamiltonian symplectomorphisms have not been made. 
On the contrary, there has been a substantial effort invested toward such constructions, starting with  \cite{Vit, Schw, Oh} and continuing to this day.\\

 A different way of looking at this issue is by not trying to explicitly construct increasingly general norms on the space of Hamiltonian symplectomorphisms, 
 and in investigating their properties individually, but in trying to establish as general properties as we can, only constrained by the general properties of a norm. 
 Moreover, and for reasons of elegance and simplicity, we are interested in norms that are bi-invariant under the action of \ $\mathsf{Ham}(\mathcal{M}, \omega)$. \
 This level of general considerations, allows far too many possible norms to become part of the discussion. To get the wealth of such norms  under control,
  and be able to make any non-trivial statements, one has to impose a relatively weak equivalence relation. It turns out that the well-known norm-equivalence 
  is appropriate for such purposes.  We recall that two norms \ $\| \cdot \|_1$ \ and \ $\| \cdot \|_2$ \ on a normed vector space, finite of infinite dimensional, are equivalent, 
  if there is a constant  \ $C\geq 1$ \ such that  
\begin{equation}
   \frac{1}{C} \  \| \cdot \|_2 \ \leq \ \| \cdot \|_1 \ \leq \ C \  \| \cdot \|_2
\end{equation}

To be more concrete, the question that was raised was to find the \ $\mathsf{Ham}(\mathcal{M}, \omega)$ \ invariant norms on \ $\mathsf{ham}(\mathcal{M}, \omega)$ \ which give rise to
bi-invariant metrics on \ $\mathsf{Ham}(\mathcal{M}, \omega)$. \ The answer was provided through in a couple of papers \cite{OW, BO}, and is stated here   
at a level of generality sufficient for our purposes. It  is taken from \cite{Ostr} nearly verbatim. Let \ $(\mathcal{M}, \omega)$ \ be a closed symplectic manifold. Any bi-invariant 
Finsler pseudo-metric on \ $\mathsf{Ham}(\mathcal{M}, \omega)$ \ obtained by a pseudo-norm \ $\| \cdot \|$ \ on \ $\mathsf{ham}(\mathcal{M}, \omega)$ \ which his continuous in the 
\ $C^\infty$ \ topology, is either identically zero, or equivalent to Hofer's metric. In particular, any non-degenerate bi-invariant Finsler metric on \ 
$\mathsf{Ham}(\mathcal{M}, \omega)$ \  which is generated by a norm which is continuous in the \ $C^\infty$ \ topology gives rise to the same topology on \ 
$\mathsf{Ham}(\mathcal{M}, \omega)$ \ as the one induced by Hofer's metric. \\

Two issues warrant  comments before closing this section. The first has to do with the fact that the statement of the previous paragraph is applicable only for closed manifolds. 
However, a typical phase space \ $\mathcal{M}$ \ encountered in Physics is the cotangent bundle of the space of generalized positions. As such, such a phase space  is not compact, 
hence not closed. Moreover, it is entirely possible that the space of generalized positions itself is a manifold with boundary. Hence the above statement of uniqueness does not apply. 
A first response to this is that there are phase spaces of interest in Physics which are closed manifolds, hence the above statements have their use when applied to such phase spaces.
A second, more heuristic, response is that in many cases in Physics, we tend to approximate a non-compact manifold with an exhaustive sequence of compact manifolds, or alternatively use an 
appropriate compactification of the underlying manifold to sidestep issues of non-compactness. Hence
one could apply the above uniqueness results to each member of the approximating sequence of compact manifolds and heuristically believe that in the appropriate limit the results will
still hold, if not in the greatest possible generality, at least for physically interesting cases. \\    

The second issue pertains to set-theoretic topology. We are used to working with the compact-open topology in the overwhelming majority of cases. This is a topology induced by the metric and 
appears to be the most natural, from a physicist's viewpoint. It should be noted that other topologies have also appeared occasionally in modelling physical systems
but have not yet given convincing enough physical results to warrant widespread and further investigation. There are two obvious exceptions to this statement. 
The first one are spaces endowed with indefinite metrics such as the ones (space-times) appearing in Relativity. 
The second one is when one deals with infinite dimensional manifolds, such as \ $\mathsf{Ham}(\mathcal{M}, \omega)$. \ For the infinite dimensional case, 
it is not necessarily clear what would constitute a useful topology for Physics. Things become especially acute when one has to define a Calculus in such infinite dimensional spaces    
as in \cite{KM}. A physicist can again be very heuristic, and pretend that infinite dimensional spaces behave as some form of inductive, in dimension, limits  of
finite dimensional spaces, turn a blind eye to obvious differences such as lack of local convexity, or of the exponential map not being surjective on infinite dimensional Lie groups etc, 
and proceed in obtaining physical predictions by ignoring all  infinite dimensional topological subtleties. 
However, we believe that a more careful look may be warranted at this point, as physical predictions may turn out to be dependent on the topology used. \\


\section{Quantization}

Canonical quantization can be traced back to the foundational ideas of P.A.M. Dirac, and the mathematical works of J. von Neumann, H. Weyl, M. Born and P. Jordan 
\cite{deG1, deG2}.  As is well-known,
it  tries to assign to functions on the phase space \ $f: \mathcal{M} \rightarrow \mathbb{R}$ \  self-adjoint operators \ $\widehat{f}: \mathcal{H} \rightarrow \mathcal{H}$ \ 
acting on an appropriately chosen  Hilbert space \ $\mathcal{H}$. \ This Hilbert space is, in the most simplistic realization, the space \ $L^2(\mathcal{M})$ \ 
of square integrable functions on \ $\mathcal{M}$. \ Let us assume \ $\mathcal{M} = \mathbb{R}^n \times \mathbb{R}^n$ \ for simplicity. 
This assignment has to satisfy several ``reasonable"  requirements,  such as   
\begin{enumerate}
   \item[({\sf Q1})]  The correspondence \ $f\mapsto \widehat{f}$ \ is linear. 
   \item[({\sf Q2})]   The constant and equal to one function should be mapped to the identity operator.
   \item[({\sf Q3})]  Consider any function \ $\phi: \mathbb{R} \rightarrow \mathbb{R}$ \  for which \ $\widehat{\phi\circ f}$ \ and \ $\phi (\widehat{f})$ \ are well-defined. Then 
                  \begin{equation}
                              \widehat{\phi\circ f} \ = \ \phi (\widehat{f})   \nonumber
                  \end{equation}
     \item[({\sf Q4})]   The operators \ $\widehat{q}_k, \widehat{p}_k, \ \ k = 1, \ldots,  n$  \ which are the quantum analogues of the coordinate functions 
                            \ $(q_k, p_k), \ \ k = 1, \ldots, n$ \ are given by 
                            \begin{equation}
                                  \widehat{q}_k f \ = \ q_k f, \hspace{10mm} \widehat{p}_k f \ = \ i\hbar \frac{\partial}{\partial q_k} f, \hspace{15mm}  f\in L^2(\mathbb{R}^n)  \nonumber
                            \end{equation}           
\end{enumerate} 
where \ $\hbar = h/2\pi$, \ and \ $h$ \  is Planck's constant. As is well-known by now, the Stone and von Neumann theorem states that the operators in \ $\sf (Q4)$ \ are the 
unique operators, up to unitary equivalence,  acting on a Hilbert space \ $\mathcal{H}$ \  which satisfy the following two conditions 
\begin{itemize}
       \item[({\sf Qa})] there are no other subspaces of \ $\mathcal{H}$, \ except \ $\{ 0 \}$ \ and  \ $\mathcal{H}$ \ itself, which remain invariant under the action of all  \ \ 
                                  $\widehat{q}_k, \ \widehat{p}_k, \ \ k=1,\ldots, n$        
       \item[({\sf Qb})] they satisfy the canonical commutation relations 
                    \begin{equation}
                           [\widehat{q}_j, \widehat{q}_k] \ = \ [\widehat{p}_j, \widehat{p}_k] \ = \ 0, \hspace{15mm}   [\widehat{q}_j, \widehat{p}_k] \ = \ i\hbar \  \delta_{jk} ,    
                                       \hspace{20mm} j, k = 1,\ldots, n      \nonumber
                    \end{equation}            
\end{itemize}
It turns out that one would like the canonical commutation relations \ $\sf{(Qb)}$ \ to be valid for all ``quantizable" functions \ $f: \mathcal{M} \rightarrow \mathbb{R}$, \ 
namely to have the commutation relations 
\begin{itemize}
   \item[({\sf Q5})] \hspace{20mm}  $[f_1(\widehat{q}_1, \ldots, \widehat{q}_n, \widehat{p}_1, \ldots, \widehat{p}_n), f_2(\widehat{q}_1, \ldots, \widehat{q}_n, \widehat{p}_1, \ldots, \widehat{p}_n)] 
                 \ = \ i\hbar \widehat{\{ f_1, f_2 \} }$ 
\end{itemize} 
where the curly brackets on the right-hand side stand for the Poisson brackets (2).  
This is the approach to quantization that P.A.M. Dirac has strongly advocated \cite{Dirac}. \
 The requirement \  {\sf (Q5)} \ is trivially satisfied,  if \ $f$ \ depends only on the canonical positions alone, 
namely if \ $f(q_1, \ldots, q_n)$, or if \ $f$ \ depends only on the canonical momenta alone, namely if \ $f(p_1, \ldots, p_n)$. \ 
However \ {\sf (Q5)} \  is no longer true for general functions \ $f$ \ which depend on both the canonical positions and the canonical momenta, 
except if these functions are at most linear in all of these canonical coordinates.\\

It turns out that it is impossible to impose all the conditions {\sf (Q1) - (Q5)} at the same time as they are mutually inconsistent. An early result is that of 
H.J. Groenewold \cite{Groen}, which was elaborated upon by L. van Hove \cite{vH}, and that of \cite{AB}. Since these early days considerable effort has been invested 
in sidestepping these difficulties and other difficulties that have emerged in attempts to canonically quantize a system. A multitude of approaches have been put forth 
in the framework of canonical quantization (the literature on this topic  is huge but we restrict our references only to a few of the  ones which we needed for the 
present work such as  \cite{GW, Kl1, Kl2, Kl3, deG1, deG2, AE}), but none of them has turned out to be  is a priori superior to all others and can be universally applied to all systems under study. 
This has probably motivated the dictum of L.D. Faddeev that ``quantization is an art, not a science".  Eventually it is agreement with experimental data that decides 
what is a successful quantization approach for a particular system from the multitude of mathematically consistent possibilities that have been developed. \\

A way to restrict the number of possible approaches has been to realize that  phase spaces of physical systems usually have additional structure, 
hence the dictum of $\cite{GW}$ that we rarely have to quantize a symplectic manifold without any additional structure.  
And all this takes place even before someone goes to classical/quantum field theory where the 
Stone, von Neumann theorem does not apply and one has even more choices in quantization, 
therefore even more decisions have to be made at different points regarding on how one must proceed. This wealth of 
choices has probably nowhere been more clear, to the authors' knowledge, than in the potential approaches and choices made therein, in quantizing gravity \cite{Sork, Loops, CAN}.\\ 

Given all the above comments, the question remains what can someone say for the phase space of a system at the ``crude level" of symplectic manifolds,
before proceeding with adding extra structure on such a phase space and then quantizing it. We see especially in {\sf (Q4)} that the coordinates of the phase space play a different 
role in Classical and Quantum Physics. In Classical Physics, they parametrize the phase space. Hence they provide a parametrization of the kinematic structure of the theory. 
The dynamics is provided by the symplectic structure or, in more physical terms, by the Hamiltonian of the system. By contrast, in Quantum Physics the kinematic structure is provided by the 
Hilbert space \ $\mathcal{H}$, \ which in the crudest treatment is taken to be the space of square integrable functions \  $L^2(\mathcal{M})$. \  Hence the coordinates play the role of operators
on \ $\mathcal{H}$, \ which are the quantum analogue of the (Hamiltonian) symplectomorphisms of \ $\mathcal{M}$. \  In category-theoretic language,   quantization cannot be a functor 
from the category of symplectic manifolds and symplectomorphisms to the category of Hilbert spaces and unitary operators, primarily because the analogue of the symplectic manifolds 
in the former category is not the Hilbert spaces of the latter, but appropriate self-adjoint operators acting on such Hilbert spaces. And the space of compactly supported Hamiltonian 
symplectomorphisms of \ $\mathcal{M}$ \ is mapped to the space of unitary operators \ $\mathfrak{U}(\mathcal{H})$ \  on that Hilbert space \ $\mathcal{H}$. \\

As was mentioned above, the only non-trivial (among ``relatively simple" Finslerian) norms in \ $\mathsf{Ham}_c(\mathcal{M}, \omega)$ \ is the Hofer norm. Its analogue, in the case of 
linear operators among Hilbert spaces, finite of infinite dimensional,  is the operator norm. We recall that the operator norm \ $\| \cdot \|_{op}$ \ of a linear operator 
 acting on a Hilbert space \ $\mathcal{A}: \mathcal{H} \rightarrow \mathcal{H}$ \  is defined as 
\begin{equation} 
   \| \mathcal{A} \|_{op} \ = \ \inf_{c} \{ c\geq 0: \ \  \| \mathcal{A} v \| \leq c\| v \|, \ \  \ \forall \ v\in V \}   
\end{equation}
or equivalently as
\begin{equation}
 \| \mathcal{A} \|_{op} \ = \  \sup \{  \|\mathcal{A}v \|, \ \ \forall \ v\in V, \ \ \|v \| = 1 \} 
\end{equation}
The Hofer norm is of class \ $L^\infty$ \ since it is defined by (20).  The same can be said about the operator norm (39), (40). 
The analogue of the \ $L^p, \ p: \mathrm{finite}$  norms (37), is given the by the Schatten $p$-norms, for \ $p\geq 1$  \ which are  defined as
 \begin{equation}
   \| \mathcal{A} \|_p \ = \ \left[\mathsf{Tr} \left( (A^\ast A)^\frac{p}{2} \right)      \right]^\frac{1}{p}
\end{equation}
where \ $\mathsf{Tr}$ \ stands for the trace  of tis argument, and where \ $\mathcal{A}^\ast$ \ is the adjoint of \ $\mathcal{A}$ \ with respect to the inner product of \ $\mathcal{H}$.  \ 
By definition, \ $\| \mathcal{A} \|_\infty$ \ is the operator norm \ $\|\mathcal{A}\|_{op}$, \ so one can see that 
\begin{equation} 
            \| \mathcal{A} \|_{op}  \ = \ \lim_{p\rightarrow\infty} \| \mathcal{A} \|_p
\end{equation}
It may also be worth noticing that the Schatten $p$-norms and the operator norm are left and right  invariant under the action of the unitary group \ $\mathfrak{U}(\mathcal{H})$ \ 
in complete analogy with the action of \ $\mathsf{Ham}_c (\mathcal{M}, \omega)$ \ on the $L^p$ distance (37), and the Hofer norm (20).   
It may be worth remarking that the operators \ $\widehat{q}_k, \ \widehat{p}_k, \ k=1, \dots, n$ \ appearing in quantization are unbounded, 
so one may have to either compactify the phase space \ $\mathcal{M}$, \ if it is not already compact, or assume that such operators have locally compact domains. 
This is in analogy with the action of \ $\mathsf{Ham}_c(\mathcal{M}, \omega)$ \ where compactly supported Hamiltonians are considered.  \\

Having defined norms on both \ $\mathsf{Ham}_c(\mathcal{M}, \omega)$ \  and on \ $\mathfrak{U}(\mathcal{H})$  \  one can now see the reason about the 
distinct role that the Cartesian coordinates play in quantization, as noticed by P.A.M. Dirac. 
It is indeed true that coordinates have no physical meaning but provide only labels that allow someone to perform explicit calculations or make any measurements.
At the end, all physical results should be independent of such coordinates. The same applies to the parametrization of the automorphisms of the underlying structures, 
such as \ $\mathsf{Ham}(\mathcal{M}, \omega)$ \ and \ $\mathfrak{U}(\mathcal{H})$. \ However, sooner or later some metrics invariably creep in the quantization process.
We want the structure of the classical category and that of the quantized category to be as close as possible. 
The unit ball of the Hofer norm, like that of any \ $L^\infty$ \ norm, is a parallelepiped whose faces are perpendicular to the Cartesian coordinate axes.   
All other norms are trivial on \ $\mathsf{Ham}_c(\mathcal{M}, \omega)$. \ 
Therefore, to reflect the uniqueness of the Hofer norm, one should also pick a parallelepiped as the unit ball of a metric in the space of self-adjoint operators of \ $\mathcal{H}$. \ 
But this is already done, since one uses the operator norm which is a sup-norm for measuring the action of all operators on the Hilbert space \ $\mathcal{H}$. \ 
Hence, it is this analogy between the available metric structures  on both the classical and the quantum systems, and the desire for the quantum category 
to reflect as much as possible the structure and properties of the classical category, that makes the Cartesian coordinates so special in quantization. \\   

One obvious question, given the close relationship between Classical and Quantum Physics, is what is the classical  analogue of Heisenberg's uncertainty principle 
\cite{deG3, deG4}. The answer, which is not obvious, are the symplectic capacities which measure, in a symplectic sense, the size of  symplectic manifolds \cite{HZ, McDS}. 
This realization was established  through the foundational  work of M. Gromov \cite{Gromov}. The construction,  properties and explicit calculation of such symplectic capacities 
have become an expanding research topic ever since. Of particular interest is the conjecture that all symplectic capacities coincide for convex bodies on \ $\mathbb{R}^{2n}$. \
This conjecture was established for some well-known symplectic capacities, for centrally symmetric convex bodies, when the dimension \ $n$ \ becomes very large \cite{GO}. \
The significance of this is that  centrally symmetric convex bodies are the unit balls of Finsler metrics on \ $\mathbb{R}^{2n}$, \ so choosing such metrics may asymptotically 
guide us toward unique aspects in the quantization of  systems with many degrees of freedom. Further developments on  the canonical quantization of such systems, may prove to
be especially fruitful for gaining better analytical description to behavior such as quantum chaos, and even contribute toward a better understanding of quantum phase transitions.  \\ 
         

\section{Discussion and conclusions}

In this work we have attempted to present an explanation why canonical quantization seems to work well  when the Hamiltonian and the canonical commutation relations are expressed 
in Cartesian, versus any other coordinate system. Our argument relied on the uniqueness of the Hofer distance on the space of compactly supported Hamiltonian symplectomorphisms, a
metric which is of \ $L^\infty$ \ class, when all other \ $L^p$ \ distances turn out to be trivial. Then the requirement that the structure obtained as a result of quantizing the classical system 
should reflect as closely as possible the behavior of the classical system itself leads us to the desired conclusion. \\

There has been a tremendous effort, for about a century, toward understanding the formal aspects and the physical requirements that go into quantizing a system. The importance of such a 
task is obvious, and cannot be understated from a physical viewpoint. Despite such efforts and significant advances in our understanding of such a process, the overall picture even in the better 
understood canonical quantization (which is usually contrasted, not quite correctly, to the functional integral quantization) remain still not very well understood. Further work in this direction
can benefit both the development of new mathematical approaches as well as help elucidate aspects of canonical quantization that may be helpful in quantizing in the future physically 
fundamental systems, such as gravity. Taking advantage of relatively recent advances in conceptual and technical aspects of Symplectic and Contact Geometry, such as Floer Homology,
Gromov-Witten invariants, quantum cohomology, etc may also prove to be quite fruitful more generally,  outside the realm of supersymmetric and string-related models in which
they are currently employed.\\   

We have to admit that the argument that we have presented is heuristic, at best. 
An objection that may arise is that the classical to quantum analogy is not perfect. Therefore the desiratum of the close resemblance, in a categorical sense, 
of the classical and the quantum structures does not have to hold.  Indeed, there are quantum 
systems that have no classical analogues, classical systems which are the limits of many different quantum systems, several inequivalent ways of quantizing a classical system, 
thus obtaining many quantum systems etc. And at the end of the day all this is general, convenient and even elegant mathematical formalism. It is unclear however, how much
of all this formalism can be actually used to describe systems which are realized in nature.  Therefore, this work does not  move substantially in the direction of making the 
quantization process a science rather than an art, but at least it may provide some hint on the reason why P.A.M. Dirac's footnote \cite{Dirac} expresses a valid fact. \\


              \vspace{7mm}

\noindent{\bf Acknowledgement:} One of us (N.K.), is grateful to Professor Kyriakos Papadopoulos for numerous discussions on set-theoretic topology and order theory, 
especially as they may possibly pertain to Physics. We would like to thank Professor Anastasios Bountis whose  continuous support made this work possible.



\end{document}